\documentclass[reprint, aip, amsmath,amssymb, superscriptaddress, jcp, citeautoscript]{revtex4-2}
\usepackage{graphicx}
\usepackage{dcolumn}
\usepackage{bm}
\usepackage{braket}
\usepackage{verbatim}
\usepackage{lipsum}
\usepackage{enumitem}
\usepackage{makecell}
\usepackage{relsize}
\usepackage{hyperref}
\usepackage{xr}
\usepackage[dvipsnames]{xcolor}
\usepackage{bbold}
\usepackage{cases}
\usepackage{upgreek}
\usepackage[normalem]{ulem}
\usepackage{textgreek}
\usepackage{gensymb}

\newcommand{\beq}{\begin{equation}}
\newcommand{\eeq}{\end{equation}}
\newcommand{\beqarray}{\begin{eqnarray}}
\newcommand{\eeqarray}{\end{eqnarray}}

\begin{document}

\title{ 
Numerically exact quantum dynamics with tensor networks: \\ Predicting the decoherence of interacting spin systems
}
\author{Tianchu Li}
\thanks{These two authors contributed equally}
\affiliation{Department of Chemistry, University of Colorado Boulder, Boulder, Colorado 80309, USA\looseness=-1}
\author{Pranay Venkatesh}
\thanks{These two authors contributed equally}
\affiliation{Department of Chemistry, University of Colorado Boulder, Boulder, Colorado 80309, USA\looseness=-1}
\author{Nanako Shitara}
\homepage{Present address: Chemical Physics Theory Group, Department of Chemistry, University of Toronto, Ontario M5S 3H6, Canada}
\affiliation{Department of Chemistry, University of Colorado Boulder, Boulder, Colorado 80309, USA\looseness=-1}
\affiliation{Department of Physics, University of Colorado Boulder, Boulder, Colorado 80309, USA\looseness=-1}
\author{Andr\'es Montoya-Castillo}
\homepage{Andres.MontoyaCastillo@colorado.edu}
\affiliation{Department of Chemistry, University of Colorado Boulder, Boulder, Colorado 80309, USA\looseness=-1}
\date{\today}

\begin{abstract}
Predicting the quantum dynamics of promising solid-state and molecular quantum technology candidates remains a formidable challenge. Yet, accessing these dynamics is key to understanding and controlling decoherence mechanisms---a prerequisite for designing better qubits, sensors, and memories. We leverage a matrix product state representation to introduce a numerically exact and scalable method to achieve this goal. We demonstrate that our method accurately predicts coherence and population dynamics of spin networks across a wide range of parameter regimes, encompassing nuclear spin sensors and qubits in solid-state semiconductors and molecular magnets. Our method further predicts spin dynamics under the influence of repeated light pulses, which are commonly used to mitigate decoherence and perform quantum sensing experiments. Our method thus provides reliable results for moderately-sized spin platforms spanning molecular magnets and solid-state spins that can guide the development of approximate but efficient quantum dynamics methods and enable principled inquiry into decoherence mechanisms.
\end{abstract}

\maketitle


The long-lived coherence of spins in many solid-state systems~\cite{awschalom2018quantum} and the versatility of molecular platforms~\cite{graham2017forging} make them promising platforms for qubits, quantum memories, and sensors. To guide their design\cite{DegenRevModPhys2017,rondin2014magnetometry,balasubramanian2008nanoscale}, one needs the ability to identify which environmental interactions affect a spin's coherence dynamics. Recent studies \cite{Jackson2022JPCC, martinez2023impact, mullin2023quantum} have documented how coherence times depend on applied magnetic fields, temperature, and solvation environments (e.g., ligand structure around molecular magnets). However, precise quantum control and sensing necessitate moving beyond broad decay timescales and zeroing in on microscopic insights from unique dynamical signatures. Achieving this requires accurate methods to simulate the coherence dynamics in many-spin Hamiltonians, as these encode a sensor's response to its detailed microscopic environment. 

Spin Hamiltonians featuring general sensor-bath and intra-bath interactions describe promising platforms for quantum technology applications, including nitrogen vacancy defects in diamond~\cite{wang2013spin,awschalom2018quantum}, nuclear spin qubits in silicon~\cite{burkard2023semiconductor, dreher2012nuclear}, and molecular magnets~\cite{sessoli1993magnetic,wasielewski2020exploiting,gaita2019molecular}. In many systems of interest, intra-bath interactions are weak, motivating the adoption of spin-star type models \cite{prokof2000theory, breuer2004non} that neglect these interactions. In such cases, one can already simulate large systems~\cite{lindoy2018simple, lewis2014asymmetric, lewis2016efficient}. However, this simplification fails when simulating commonly employed measurements obtained from the application of multiple pulse sequences, engendering \textit{unphysical} coherence dynamics that recur indefinitely~\cite{Rowan1965, Zhao2012, Maze2008PRB, Szankowski2017JP}. 

Over the past two decades, the cluster-correlation expansion (CCE) method~\cite{Yang2008CCE1, Yang2009CCE2} has emerged as a leading tool to tackle this problem. It has successfully reproduced coherence decay timescales consistent with experimental observations in various solid-state~\cite{Yang2017ProgPhys, Ma2014natcomm, Balian2014PRB, Seo2016natcomm, Bourassa2020natmat, Onizhuk2021PRXQ} and molecular~\cite{jeschke2025MR, Jahn2024JCP, Bahrenberg2021MR, Canarie2020JPCL, Kveder2019JCP} systems. However, CCE suffers from significant challenges that limit its broad applicability. For example, it does not converge uniformly with expansion order or bath size, and the \textit{multiplicative} nature of its cluster-based expansion causes numerical instabilities. Thus, CCE becomes inaccurate in many physically relevant parameter regimes, such as in strongly interacting spin baths \cite{Witzel2012, onizhuk2024PRL, schatzle2024spin}. While CCE can nevertheless obtain order-of-magnitude estimates for decoherence times, developing next-generation quantum technologies requires a \textit{precise} understanding of the dissipative processes that assail quantum systems and their signatures in coherence measurements.

\begin{figure*}[ht!]
    \includegraphics[width=0.9\linewidth]{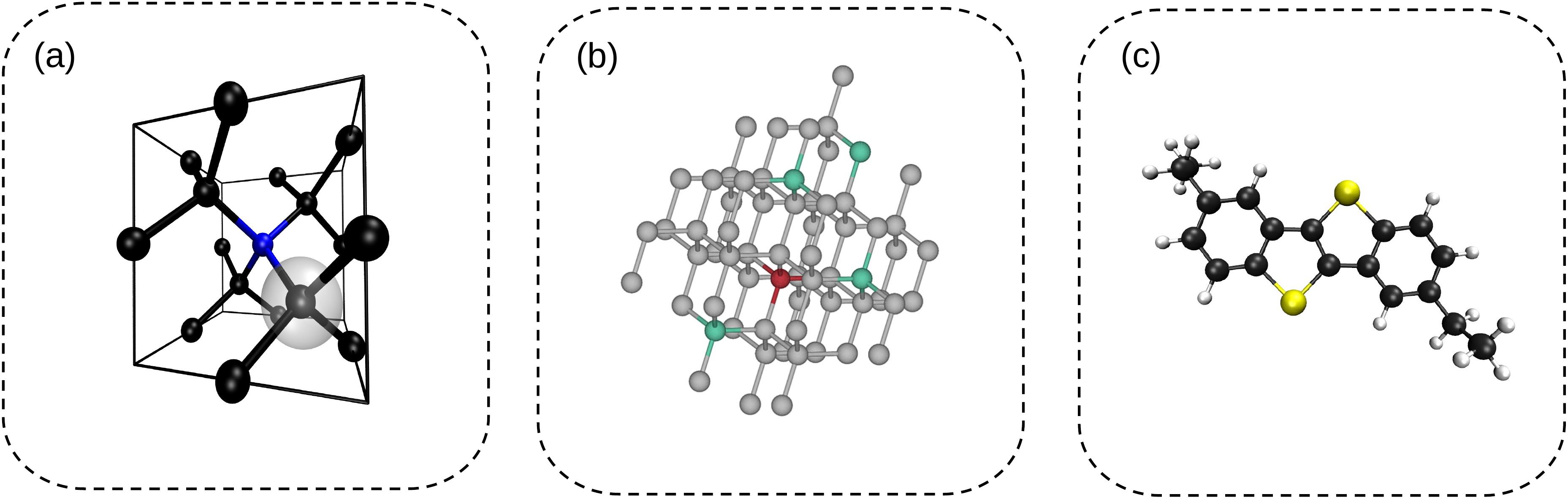}
    \caption{(a) BSBS molecule with ethyl substituents (BSBS-2Et). (b) $^{31}$P defect (red) in a Si lattice, with $^{29}$Si (green) and all other Si isotopes (gray). (c) NV defect in diamond, with nitrogen (blue) and the vacancy (white).}
    \label{fig:schematics}
\end{figure*} 

Here, we develop and demonstrate the accuracy and broad applicability of our Spin Bath-Truncated MPS (SB-tMPS) method, which leverages a matrix product state (MPS) representation to simulate the dynamics of general interacting spin-bath Hamiltonians. However, growing entanglement generally raises the cost of the MPS~\cite{verstraete2008matrix, Schollwock2011AP, orus2014, qin2022hubbard}, making simulations of interacting spin networks challenging. We tame this cost via a judiciously chosen low-rank MPS representation. For highly entangled cases, our GPU-accelerated MPS implementation enables efficient simulations, allowing us to simulate systems of up to $\sim$100 spins within a few hours on a NVIDIA V100 GPU. This cost would be at least tenfold larger on a state-of-the-art CPU, as the GPU's massively parallel architecture and high memory bandwidth are ideally suited to the tensor contractions and matrix decompositions that dominate our computational workload. Further, our SB-tMPS accesses arbitrary observables and accurate long-time dynamics, and treats mixed spin species (e.g., $S=1/2, 1, 3/2$). Our SB-tMPS can also accurately predict dynamics under \textit{arbitrary} control pulse sequences commonly employed in experiments (see Supplementary~Information~(SI)~Sec.~V). Unlike CCE, our SB-tMPS consistently achieves numerical convergence with simulation parameters. We illustrate our method's applicability to and accuracy across realistic systems where the CCE method becomes unstable: a $^{31}$P defect in Si that functions as a solid-state nuclear spin qubit, and a derivative of [1]Benzoseleno[3,2-b]benzoselenophene (BSBS) molecule that has attracted interest for information storage and spintronics~\cite{naber2007organic, dediu2009spin, sanvito2007spintronics, lunghi2020insights}. Our results suggest that our SB-tMPS substantially extends the applicability of numerical quantum dynamics simulations for experimentally relevant spin-bath systems, surpassing the limitations of CCE.


\begin{figure*}[t]
    \includegraphics[width=\textwidth]{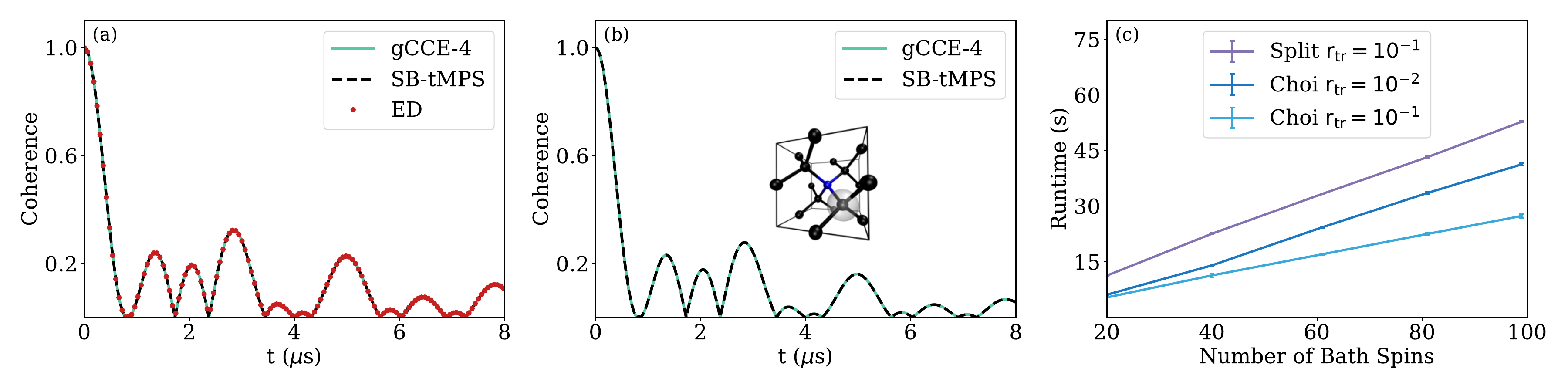}
    
    \caption{
    Coherence dynamics for a representative NV center sensor. (a) Free induction decay dynamics obtained for an NV center with a spin bath consisting of 5 $^{13}$C and 1 $^{14}$N spins using the gCCE-4, SB-tMPS, and ED methods, showing visual agreement. (b) Free induction decay dynamics obtained for the same NV center with a spin bath consisting of 98 $^{13}$C and 1 $^{14}$N spins. The agreement between CCE and SB-tMPS results persists. Inset: structure of NV center in diamond. (c) Scaling of 10 steps computational time using different splitting methods (Choi or Split), and using different SVD truncation thresholds.}
    \label{fig:NV}
\end{figure*}

Our SB-tMPS treats general Hamiltonians describing a central spin (sensor) coupled to an interacting spin bath (see SI~Sec.~I). This Hamiltonian can include self (zero-field splitting for electron spins and quadrupole interactions for nuclear spins) and mutual interaction terms. Unlike spin impurity problems, we also include intra-bath interactions as these mediate the experimentally observed decay of coherence dynamics under applied pulse sequences. In SI~Sec.~II, we parameterize this Hamiltonian with standard electronic structure methods~\cite{neese2020orca,perdew1996generalized,grimme2010consistent,adamo1999toward}. 

Our SB-tMPS evolves the many-spin density matrix by adopting the MPS formulation~\cite{Schollwock2011AP, Orus2014AP} while combining innovations that tame the computational costs of tackling realistic systems (see SI~Sec.~III). First, while it is common to use the ``split'' MPS representation~\cite{guan2024mpsqd, Shi2018JCP, ke2022hierarchical, Jaschke2018QST}, we vectorize the Hilbert space with the Choi transformation~\cite{li2023tangent} (shown in Fig.~S1(b)), enabling us to construct matrix product operators with a lower bond dimension and reducing the method's computational cost. Second, we employ the time-dependent variational principle (TDVP)~\cite{Haegeman2016PRB} with a tangent space projection~\cite{haegeman2011time, lubich2015time, Haegeman2016PRB, yang2020time, hackl2020geometry}, fixing the bond dimension of the MPS. In contrast, the common practice of using a traditional differential equation solver to propagate the MPS chain\cite{oseledets2011tensor} requires a truncation step to ameliorate but not fix the bond dimension growth\cite{greene2017tensor, guan2024mpsqd}, rendering long-time dynamics inaccessible. Finally, since bond dimension grows with increasing connectedness of the Hamiltonian, we leverage the hierarchy of coupling strengths in spin Hamiltonians to inform singular value decomposition (SVD) thresholds and maintain computational costs low. We employ two different SVD thresholds, $r_{tr}$, based on the ratio of the sensor-bath ($\lambda_{sb}$) and intra-bath ($\lambda_{bb}$) couplings: $\lambda_{sb}/\lambda_{bb}$. When $\lambda_{sb}/\lambda_{bb} \gg 1$, we truncate the singular values corresponding to intra-bath interactions aggressively, using a larger SVD threshold. For example, in NV centers, where $\lambda_{sb}/\lambda_{bb} \sim 10^{3}$, we employ $r_{tr}^{sb} = 10^{-14}$ for sensor, bath, and sensor-bath MPOs, and a more aggressive $r_{tr}^{bb} = 10^{-1}$ for intra-bath MPOs. This strategy---which can be used to truncate other weak-coupling terms, e.g., long-range interactions---tames the growth of the MPO bond dimension, enabling the simulation of many ($\sim200$) bath spins, while maintaining high accuracy.


We can now assess our method's performance in predicting the coherence dynamics of paradigmatic examples of quantum sensors: an NV center, a $^{31}$P defect in Si, and a derivative of the BSBS molecule with ethyl substituents. Beyond showing individual promise as quantum sensors, the Hamiltonians for these systems cover a wide swath of parameter space, posing diverse challenges to our SB-tMPS. For example, as emblematic platforms on which CCE works well~\cite{Onizhuk2023PRB, Zhang2020PRB}, NV centers lie in the weak intra-bath coupling limit. Conversely, solid-state nuclear spin qubit systems exhibit intermediate strength intra-bath coupling and are known to pose challenges for CCE. Finally, there is growing interest in CCE-based studies of molecules~\cite{jeschke2025MR, Jahn2024JCP, Bahrenberg2021MR, Canarie2020JPCL, Kveder2019JCP}---systems that often lie beyond the pure dephasing limit, where the accuracy of CCE has not been assessed. Hence, a comparison of CCE and SB-tMPS in evaluating molecular coherence would inform when CCE can be reliably applied to such systems. 

The NV center is an electronic spin defect consisting of a lattice vacancy neighboring a substituted nitrogen in a diamond lattice. Because of its almost millisecond-long coherence times~\cite{stanwix2010coherence}, even at room temperature, this platform has supported various implementations of quantum sensing~\cite{maze2008nanoscale, dolde2011electric,liu2019coherent} and networks~\cite{nemoto2016photonic, DegenRevModPhys2017}. We employ this system to illustrate the reliability of our SB-tMPS against exact diagonalization (ED) and CCE results, and demonstrate that its performance is comparable to that of CCE for systems where CCE works well. We adopt NV Hamiltonian parameters under an applied magnetic field value of $B=100$G, placing the dynamics in the pure dephasing limit. To compute the CCE results, we use the gCCE implementation in the pyCCE package~\cite{yang2020longitudinal, Onizhuk2021pycce} (see SI~Sec.~IV).

Figure~\ref{fig:NV}~(a) compares the Ramsey coherence measurement of an NV center surrounded by a bath of five~$^{13}$C and one~$^{14}$N (spin-1) nuclear spins, with dynamics calculated using CCE, SB-tMPS, and ED. Henceforth, gCCE-$N$ denotes gCCE taken to order $N$. All results agree with each other, indicating that both CCE and SB-tMPS yield accurate dynamics for this system, as expected. We extend this setup to include a total of 99 bath spins and show the Ramsey coherence dynamics in Fig.~\ref{fig:NV}~(b). Both CCE and SB-tMPS results remain in agreement and exhibit no numerical instabilities, suggesting that both methods recover the correct dynamics even for moderate spin bath sizes. Given the success of both methods in this paradigmatic example, in Fig.~\ref{fig:NV}~(c), we compare the computational scaling with bath size for different choices of MPS representation and various SVD truncation thresholds. Under GPU acceleration, the length of the MPS chain determines the computational time. By reducing the MPS chain length by half, the Choi representation achieves a twofold acceleration over the split representation. The SVD truncation converges with a relatively large threshold ($r_{\rm tr} = 10^{-1}$) for NV centers and provides a promising acceleration for simulations with a large number of bath spins. Moreover, under aggressive SVD truncation, we observe a linear scaling of computational time with the number of bath spins, as seen in Fig.~\ref{fig:NV}(c), demonstrating that our SB-tMPS method can treat large systems with realistic complexity.

Having established that CCE and SB-tMPS perform well in cases with stronger sensor-bath than intra-bath interactions, we turn to the more challenging $^{31}$P donor nuclear spin in silicon---an example of a semiconducting system where the sensor is a nuclear spin and the sensor-bath and intra-bath interactions are comparable. This system holds promise as a quantum memory and can be easily integrated with existing semiconductor technology used for classical computation \cite{kane1998silicon,morton2008solid,dreher2012nuclear}. Previous attempts to study this system with CCE~\cite{Witzel2012} have achieved qualitative agreement with experimentally determined decoherence timescales, albeit only after modifying the method. This is because CCE is known to struggle in nuclear spin sensors in semiconductors where the separation of sensor-bath and intra-bath interaction strengths disappears~\cite{Witzel2012}. Below, we test the comparative performance of our SB-tMPS in this more challenging system. 

\begin{figure}[ht!]
    \centering
    \includegraphics[width=0.4\textwidth]{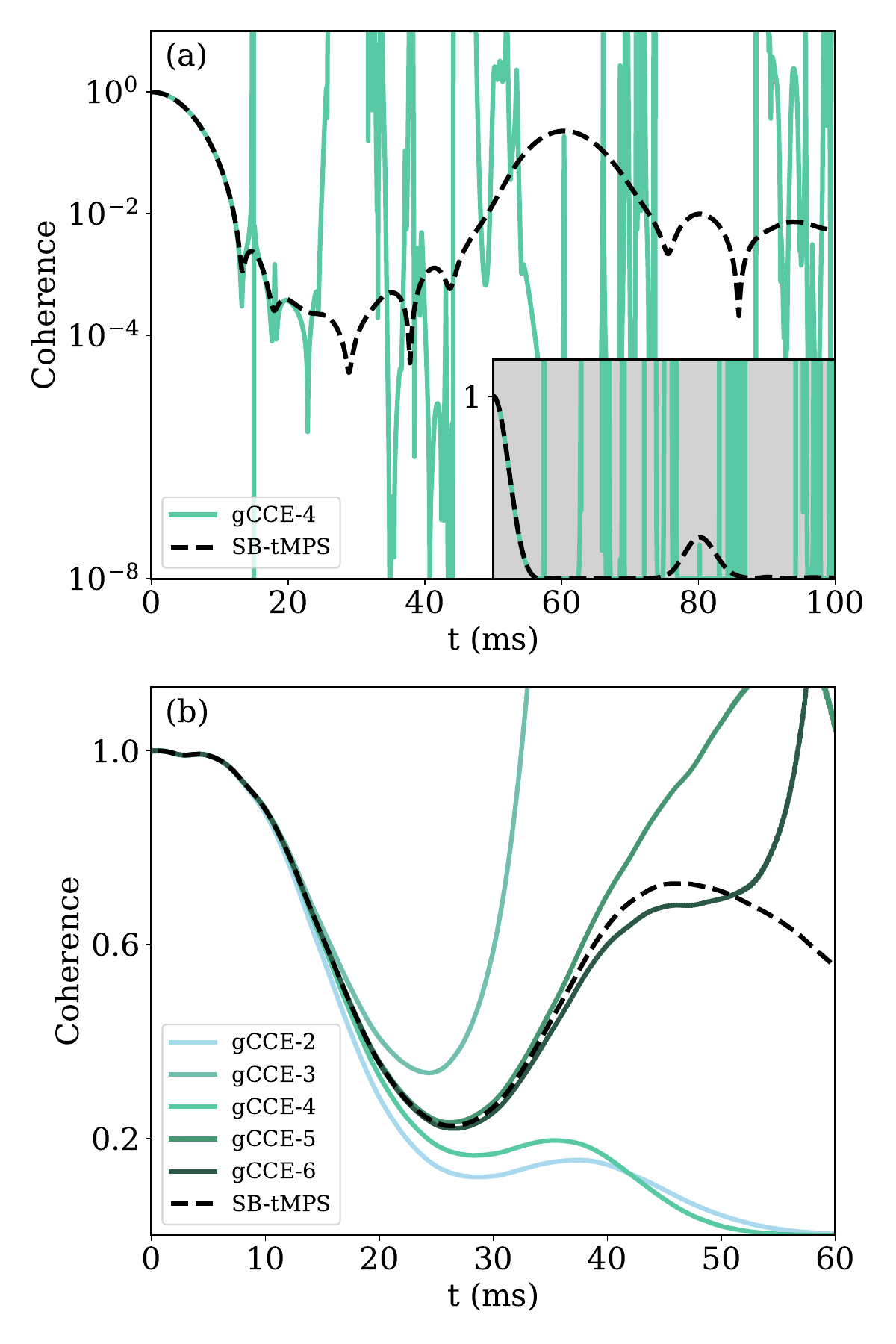}
    \caption{
    Coherence dynamics for a $^{31}$P nuclear spin qubit in a Si environment. (a) Free induction decay dynamics obtained using the gCCE-4 method (teal) and SB-tMPS (black) on a semilog scale. Inset: the same plot, on a linear scale. While the CCE dynamics show numerically unstable behavior, especially at long times, the SB-tMPS results remain stable throughout the considered evolution time. Additionally, the SB-tMPS approach can capture the finer structure in the dynamics beyond $t\sim15$~ms, unlike the CCE method. (b) Spin-Echo dynamics obtained at various CCE orders. The gCCE results show accurate dynamics at early times, but CCE dynamics at various orders differ from each other, indicating poor convergence, with results from orders $3-6$ becoming numerically unstable at longer times.
    }
    \label{fig:nuclear}
\end{figure}

Figure~\ref{fig:nuclear}~(a) compares SB-tMPS and gCCE-4 calculations of the coherence dynamics of the $^{31}$P nuclear spin in silicon under a Ramsey measurement, with the 12 nearest bath spins considered. The CCE dynamics exhibits instabilities soon after the coherence decays below 0.01, at $t\approx 15$~ms, which continue to appear more frequently at later times. Although the coherence profile beyond $t\approx 15$~ms is hard to discern on a linear scale (see inset), the SB-tMPS reveals a wealth of fine structure that becomes apparent on a logarithmic scale. In contrast, the CCE dynamics exhibit frequent divergences, indicating the method's failure at evaluating long-time dynamics.

While this dubious CCE performance can be attributed to lack of convergence with order number, we now show that the method's convergence is neither uniform nor easily affordable. Figure~\ref{fig:nuclear}~(b) shows CCE dynamics obtained with increasing cluster order for the same $^{31}$P nuclear spin system under a spin echo measurement. It demonstrates that increasing the cluster order does not strictly improve the dynamics. In fact, for cluster orders $3, 5, 6$, the coherence values exceed 1, indicating unphysical pathologies. When focusing only on the even cluster orders, the difference between the highest order, gCCE-6, and the gCCE-4 result is larger than the difference between gCCE-2 and gCCE-4, indicating the improvement in the results with increasing order is nonmonotonic. Nevertheless, results from all cluster orders agree up to $t\approx 20$~ms, consistent with the expectation that the CCE should be short-time accurate. However, this timescale is much shorter than the apparent decay time, making estimates of dephasing ($T^*$) times uncertain at best and deceptive at worst. Furthermore, in such cases, converging against cluster order becomes prohibitively expensive due to the combinatorially bounded cost of CCE with system size, casting doubt on the accuracy of the accessible CCE dynamics.


We now turn to a promising emerging technology in quantum information science: molecules. The synthetic tunability and coherence times on millisecond timescales of electron spins in molecular qubits~\cite{zadrozny2015millisecond} make them attractive candidates for quantum technologies. Recent works have started to leverage CCE to characterize the coherence dynamics of molecular systems~\cite{jeschke2025MR, Jahn2024JCP, Bahrenberg2021MR, Canarie2020JPCL, Kveder2019JCP}. However, the accuracy CCE can offer for such systems remains unclear, especially as they often lie beyond the pure-dephasing regime. Having demonstrated that our SB-tMPS accurately predicts coherence dynamics for solid-state systems, even when commonly employed convergence criteria suggest convergence of the CCE while yielding incorrect dynamics, we anticipate that our approach can offer the necessary accuracy to interrogate the coherence dynamics of molecular qubits. As an example, we consider the coherence and population dynamics of a selenophene derivative, BSBS, with ethyl substituents, which we refer to as BSBS-2Et (see Fig.~\ref{fig:schematics}(a)). The magnetic properties of BSBS-2Et have been characterized experimentally~\cite{schott2017tuning} and theoretically\cite{lunghi2020insights}, and similar thiophene and selenophene derivatives have shown promise as candidate materials for spin transport and data storage applications\cite{naber2007organic, dediu2009spin, sanvito2007spintronics}. The bath for BSBS-2Et consists of 18~spins, 16~$^1$H and 2~$^{77}$Se nuclei.

\begin{figure*}[ht!]
    \centering
    \includegraphics[width=\textwidth]{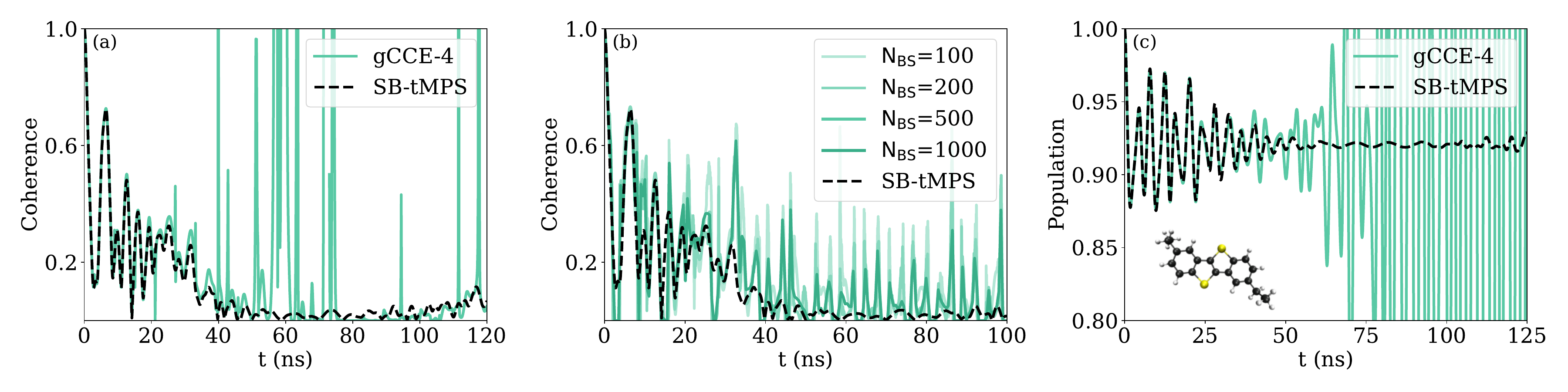}
    \caption{
    Coherence and population dynamics for the BSBS-2Et molecule. (a) Free induction decay dynamics obtained using the gCCE-4 method (teal) and SB-tMPS (black). While the CCE dynamics show numerically unstable behavior, especially at long times, the SB-tMPS results remain stable throughout the considered evolution time. (b) Free induction decay dynamics obtained with bath sampling. Bath sampling reduces numerical instabilities but results in inaccurate dynamics, even at early times. (c) Population dynamics obtained using the gCCE-4 method (teal) and SB-tMPS (black). The CCE result shows increasingly uncontrolled oscillatory behavior at later times. Inset: structure of BSBS-2Et molecule.}
    \label{fig:molecule}
\end{figure*}

Figure~\ref{fig:molecule}~(a) compares the gCCE-4 and SB-tMPS coherence dynamics of BSBS-2Et under a Ramsey measurement. The overall dynamics of coherence decay from both methods largely agree up to $\approx40$~ns, although numerical instabilities start to arise in the CCE result even before this point. Starting around 20~ns and especially beyond 40~ns, the CCE results exhibit sharp peaks and dips, some numerically divergent. In the absence of a numerically exact benchmark, it is difficult to distinguish which of these features are physical, underscoring the importance of having an independent method to verify the accuracy of the CCE dynamics. The SB-tMPS results show no such behavior, sustaining physically reasonable dynamics throughout the entire simulation time, indicating that such features in the CCE dynamics are, indeed, artifacts. We find that these instabilities in CCE dynamics arise from population relaxation effects that become relevant at low magnetic fields. In SI~Sec.~IV, we show this by recovering stable CCE results upon increasing the magnetic field or limiting the Hamiltonian to pure dephasing interactions.

Monte Carlo sampling of pure bath states has been proposed to attenuate numerical instabilities in CCE~\cite{Witzel2012, Onizhuk2021pycce, Onizhuk2021PRXQ}. While this technique recovers the original CCE dynamics upon sampling all $\mathcal{O}((2S+1)^{N})$ configurations (for $N$ spin-$S$ particles), it assumes that only some of the pure states exhibit pathological behavior and that excluding these from the sampling returns a better result. Panel (b) bath sampling shows results for the system considered in Fig.~\ref{fig:molecule}~(a) using $N_{BS}=$100-1000 samples. Although diverging dynamics disappear in panel (b), the disagreement between SB-tMPS and CCE results starts earlier than before (at $<5$~ns), whereas significant disagreement only onsets by $\sim$20~ns in panel (a). This earlier disagreement arises even when bath sampling is expected to perform optimally, i.e., $N_{BS} \rightarrow \infty$. Thus, while bath sampling may suppress numerical instabilities in CCE simulations, {\it it constitutes an uncontrolled approximation that can lead to inaccurate dynamics}. Furthermore, bath sampling increases CCE's computational cost, nullifying its efficiency. SI~Sec.~IV details our convergence tests of this system with, e.g., maximum intra-cluster spin distance, and Fig.~S9 shows that our SB-tMPS method can simulate dynamics subject to multiple applied pulses.

While coherence dynamics fully characterize decoherence in pure dephasing systems, decoherence can also happen through population relaxation. As implemented in pyCCE, gCCE can also predict population dynamics. Figure~\ref{fig:molecule}~(c) compares the population dynamics of the BSBS-2Et obtained using the SB-tMPS and gCCE-4 methods. Beyond $t\approx40$~ns, the CCE dynamics shows increasingly strong oscillations, eventually reaching unphysical values exceeding 1. In contrast, the SB-tMPS results show damped dynamics that never exceed 1. Unlike in the case of coherence dynamics, where discontinuous divergences signaled the breakdown of CCE dynamics, it is more difficult to assess when the CCE population dynamics degrade in quality in the absence of an exact benchmark. Nevertheless, the unphysical oscillations above 1 indicate a breakdown of the method's ability to predict accurate dynamics. Further, because population and coherence dynamics are coupled, divergences in the coherence dynamics can serve as an indication of the degradation in accuracy of the population dynamics, suggesting an onset of inaccurate behavior starting at $t\approx40$~ns. In contrast, our SB-tMPS provides an accurate prediction for population dynamics beyond the pure dephasing limit.

Despite the accuracy and broad applicability of our SB-tMPS, CCE still offers greater computational efficiency. Thus, the choice between methods reduces to the balance of accuracy and computational efficiency. To enable an informed choice of which method to pursue, we compare the computational costs across the two methods. CCE is efficient when the required cluster order is kept low ($\lesssim 6$), even when the bath contains many spins ($\sim 100$s), but it quickly becomes intractable at higher orders, with Ref.~\cite{jeschke2025MR} reporting a computational time scaling of $(2^c)^{2.5}$ with cluster order $c$. We show gCCE runtime scaling with cluster order and maximum cluster spin distance in SI~Sec.~VI. Moreover, neither its convergence nor accuracy is guaranteed, although the method can often identify overall decay timescales within an order of magnitude. In contrast, SB-tMPS incurs a larger computational prefactor but can be made to scale quadratically or linearly, depending on the hierarchy of interaction strengths. For weak bath interactions, such as in the NV center, SB-tMPS exhibits linear scaling with system size (see Fig~\ref{fig:NV}(c)), enabling simulations with bath sizes of up to several hundred spins. For systems with intermediate intra-bath couplings, as in $^{31}$P, SB-tMPS has power-law scaling (see SI Sec.~VI), limiting simulations to fewer than $\sim$50 bath spins. Nevertheless, SB-tMPS demonstrates significantly better numerical accuracy and stability for a wider range of Hamiltonian parameterizations. Further, one can reduce the cost of CCE by using SB-tMPS as a propagation scheme for large clusters.

Hence, CCE is best suited for cases where intra-bath interactions are sufficiently weak {\it and} when only a rough characterization of the decoherence times is required. When one needs access to \textit{accurate} dynamics and for systems with an intermediate number of environmental spins where the simplifications associated with the central limit theorem cannot be invoked~\cite{mukamel1995, Mukamel1985}, SB-tMPS provides a distinct advantage. This limit may prove to be a sweet spot for quantum sensing where signals arise from complex systems and still encode information about the many-body interactions that gave rise to them. The regime of large numbers of strongly interacting environmental spins remains difficult to address with either of these methods, with SB-tMPS offering the advantage that, given sufficient computational time and resources, it is expected to provide a more reliable result compared to CCE.


We have thus introduced SB-tMPS, an MPS-based method to simulate the dynamics of interacting spin networks subject to an arbitrary number of light pulses that describe solid-state and molecular systems of interest to quantum technology applications. SB-tMPS provides greater numerical stability and accuracy than CCE, the current state-of-the-art approach, enabling rigorous inquiry into spin-interaction-mediated decoherence pathways across near-term quantum platforms. As the need to identify sources of decoherence in spin platforms grows, accuracy becomes essential for any quantum dynamics method. Thus, the high accuracy and versatility of our SB-tMPS render it an effective tool for principled studies of decoherence dynamics and mitigation, and for guiding the development of scalable and approximate dynamics methods.

\section*{Acknowledgements}

The authors acknowledge funding from the National Science Foundation (Grant No.~2412615). A.M.C.~also acknowledges the support from a David and Lucile Packard Fellowship for Science and Engineering. The authors thank Anthony J Dominic III and Srijan Bhattacharyya for comments on the manuscript. We also thank Prof.~Qiang Shi for access to his GPU cluster while benchmarking our SB-tMPS. This work utilized the Alpine high-performance computing resource at the University of Colorado Boulder. Alpine is jointly funded by the University of Colorado Boulder, the University of Colorado Anschutz, Colorado State University, and the National Science Foundation (Award No. 2201538).

\section*{References}

\bibliography{main}

\end{document}



\setcounter{section}{0}
\setcounter{equation}{0}
\setcounter{figure}{0}
\setcounter{table}{0}
\setcounter{page}{1}

\renewcommand{\theequation}{S\arabic{equation}}
\renewcommand{\thefigure}{S\arabic{figure}}
\renewcommand{\thepage}{S\arabic{page}}
\renewcommand{\bibnumfmt}[1]{$^{\mathrm{S#1}}$}
\renewcommand{\citenumfont}[1]{S#1}

\title{Supplementary Information for \\``Numerically exact quantum dynamics with tensor networks:
Predicting the decoherence of interacting spin systems''}

\author{Tianchu Li}
\thanks{These two authors contributed equally}
\affiliation{Department of Chemistry, University of Colorado Boulder, Colorado 80309, USA}
\author{Pranay Venkatesh}
\thanks{These two authors contributed equally}
\affiliation{Department of Chemistry, University of Colorado Boulder, Colorado 80309, USA}
\author{Nanako Shitara}
\affiliation{Department of Chemistry, University of Colorado Boulder, Colorado 80309, USA}
\affiliation{Department of Physics, University of Colorado Boulder, Colorado 80309, USA}
\author{Andr\'es Montoya-Castillo}
\homepage{Andres.MontoyaCastillo@colorado.edu}
\affiliation{Department of Chemistry, University of Colorado Boulder, Colorado 80309, USA}

\maketitle

\renewcommand{\theequation}{S\arabic{equation}}
\renewcommand{\thefigure}{S\arabic{figure}}
\renewcommand{\thepage}{S\arabic{page}}
\renewcommand{\bibnumfmt}[1]{$^{\mathrm{S#1}}$}
\renewcommand{\citenumfont}[1]{S#1}

\onecolumngrid

\tableofcontents
\newpage

\section{Spin-Bath Hamiltonian and observables}

\begin{figure}[!htb]
    \centering
    \includegraphics[width=0.8\linewidth]{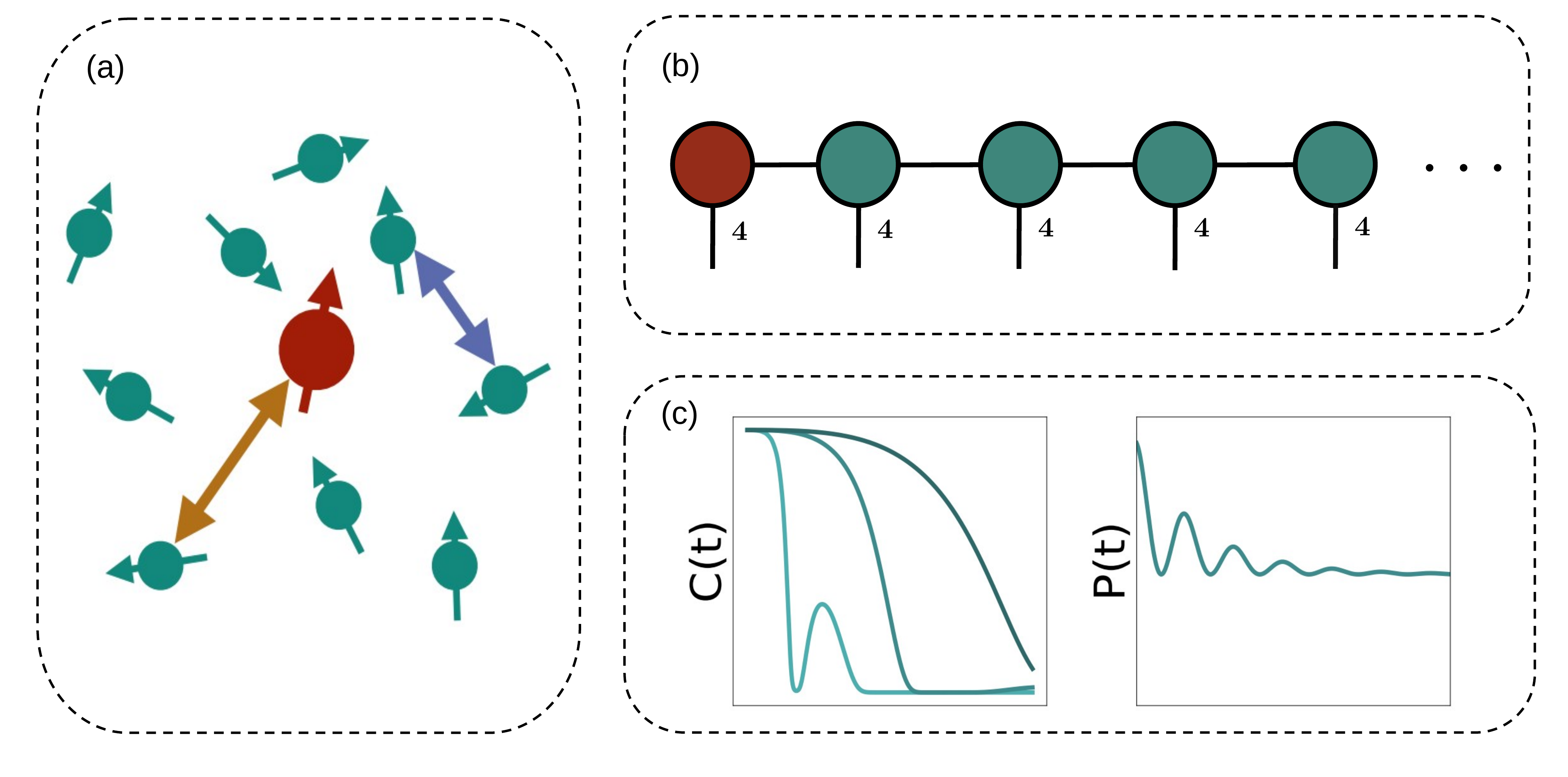}
    \caption{(a) Schematic representation of the interacting spin-bath model.(b) MPS representation in SB-tMPS. (c) Schematic output of SB-tMPS method: coherence dynamics (left) under applied pulses, and population dynamics (right).}
    \label{fig:mps-schem-SI}
\end{figure} 

Our SB-tMPS can treat a general Hamiltonian for a central spin (sensor) coupled to an interacting spin bath: 
\begin{align}
 \label{eq:ham}
    H_{\rm T} = H_{\rm s} + H_{\rm b} + H_{\rm s-b} + H_{\rm I}.
\end{align}

The sensor (\textit{s}) and bath (\textit{b}) Hamiltonians are defined as,
\begin{align}
    H_{\rm s} &= \omega_s S_0^z + \sum_{k,l\in\{x,y,z\}}S_0^k \mathbf{D}^{k,l} S_0^l \label{eq:Hs} ,\\
     H_{\rm b} &= \sum_{i=1}^K\omega_iS_i^z + \sum_i \sum_{k,l\in\{x,y,z\}}S_i^k \mathbf{P}_i^{k,l} S_i^l \label{eq:Hb}.
\end{align}
where $S_0^z$ ($S_i^z$) is the $z$ spin operator for the central ($i$th bath) spin, and $\omega_s = \gamma_s B_z$ ($\omega_i = {\bf \gamma}_i B_z$) is the central ($i$th bath) spin's frequency, where $\gamma_s$ ($\gamma_i$) is the gyromagnetic ratio of the central ($i$th bath) spin and $B_z$ is the magnitude of the magnetic field applied in the $z$ direction. $K$ is the total number of bath spins considered. The second terms in Eqs.~\ref{eq:Hs} and \ref{eq:Hb} correspond to the zero-field splitting for an electronic spin and quadrupole interaction for nuclear spins, respectively. 

The sensor-bath coupling, 
\begin{align}
\label{eq:sbham}
     H_{\rm s-b} = \sum_{i=1}^K \sum_{k,l\in\{x,y,z\}} S_0^{k} \mathbf{A}_{i}^{k, l} {S_i}^{l},
\end{align}
involves the coupling tensor, ${\bf A_i}$, between the sensor and the $i$th bath spin. In the pure dephasing limit, only $k=z$ is included. The intra-bath contribution, 
\begin{align}
    H_{I} = \sum_{i<j}\sum_{k,l\in\{x,y,z\}} S_i^k \mathbf{J}_{i,j}^{k,l}S_j^l,
\end{align}
contains interparticle interactions, where $\mathbf{J}_{i,j}$ denotes the coupling tensor connecting the $i$th and $j$th bath spins. This model is depicted schematically in Fig.~\ref{fig:mps-schem-SI} (a). These interactions are crucial as they mediate the experimentally observed decay of coherence dynamics under applied pulse sequences. We parameterize intra-bath interactions from standard geometries of the physical systems\cite{neese2020orca, perdew1996generalized, grimme2010consistent, adamo1999toward} (see SI~Sec.~\ref{SI:parametrization}).

In all our simulations, we are interested in computing experimental observables such as the coherence and population dynamics of the system sensor. The coherence and population dynamics take the following forms, respectively
\begin{subequations}
\begin{align}
   C(t) &= \frac{\left\lvert\mathrm{Tr}[(S_0^x - iS_o^y)\rho(t)]\right\lvert}{\left\lvert\mathrm{Tr}[(S_0^x - iS_o^y)\rho(0)]\right\lvert}    \label{eq:observables},\\
   P(t) &= \mathrm{Tr}[(S^z_0+1/2 \times{\mathbb{I}})\rho(t)],
\end{align}
\end{subequations}
where $\rho(t) = U^{\dagger}(t)\rho(0)U(t)$, $U(t)$ is the forward propagator, and $\rho(0) = \rho_S \otimes \rho_B$ is the full density matrix, taken to be initially uncorrelated. This initial condition is consistent with common quantum sensing and computing protocols, where the qubit of interest is addressed to be in a particular state, say the $\lvert+\rangle$ state : $\rho_S = 1/2(|0\rangle\langle 0| + |0\rangle\langle 1|+|1\rangle\langle 0|+|1\rangle\langle 1|)$, and the environment is taken to be at thermal equilibrium. Because the energy scales of these many-spin Hamiltonians span microwave frequencies, i.e., $10^{0}-10^{5}$~Hz, and the temperature of the experiments span $T \in [10^{11}, 10^{13}]$~Hz ($T \in [5, 300]$~K), it is common to invoke the infinite temperature approximation for the bath.

Experimentally, the coherence of a qubit (Eq.~\ref{eq:observables}) is often accessed via a Ramsey measurement sequence, where the qubit is initially prepared in the $\lvert+\rangle$ state, freely evolved for a fixed time $t$, and then projected into one of the population states and measured. A series of such measurements at different values of $t$ gives access to $C(t)$. This can also be augmented by applying periodic pulse sequences during the free evolution step, which has the effect of lengthening the effective coherence time.

When considering the coherence dynamics arising upon application of dynamical decoupling sequences\cite{uhrig2007keeping}, the evolution operator is split by the light pulse interactions. For instance, in a sequence with $N$ light pulses coming in at $(t_1, t_2, ..., t_N)$, the forward propagator looks thus:
\begin{equation}
\begin{split}
    U(t) = \exp_{+}\left[{-\frac{i}{\hbar}\int_0^t ds\ H_{lm}(s)}\right]
\end{split}
\end{equation}
where the full Hamiltonian, including the light-matter interaction, takes the form:
\begin{equation}
    H_{lm}(t) = H_T+ v(t)S^0_y.
\end{equation}
$v(t) = 0$ for all times when the pulse is turned off, and in general may have a time-dependent profile while the pulse is turned on. For the $\pi$-pulses that are applied in the measurements we consider, the profile of $v(t)$ must satisfy
\begin{equation}
    \int_{\tau_i}^{\tau_f} dt\,v(t) = \pi \label{eq:integrated_pulse}
\end{equation}
for each pulse applied turned on between $t=\tau_i$ to $t=\tau_f$.
Upon taking the instantaneous pulse limit, the duration that the pulse is turned on for tends towards zero while still satisfying Eq.~\ref{eq:integrated_pulse}, and as a result one can simplify the forward propagator as
\begin{equation}
\begin{split}
    U(t) = e^{-iH (t_N-t_{N-1})} R_y(\pi) ...  R_y(\pi) e^{-iH (t_2-t_1)}  R_y(\pi) e^{-iH t_1},
\end{split}
\end{equation}
where $R_y(\theta) = e^{-i\theta S_0^y}$ are pulse rotation operators that act instantaneously.

\label{SI:ham}

\section{Hamiltonian Parametrization}

To realistically model the dynamics of physical systems, we require accurate spin-bath Hamiltonian parameters, such as the gyromagnetic ratios ($\gamma_s$) of the central spin, the coupling tensors of the central spin with the bath spins ($\mathbf{A}$), and the bath-bath coupling tensors ($\mathbf{J}$). We obtain these using electronic structure calculations.

For our molecular system, BSBS-2Et, we use \texttt{ORCA}\cite{neese2020orca} to compute its geometry and magnetic properties. We performed geometry optimization of the molecule using density functional theory (DFT) with the PBE functional\cite{perdew1996generalized}, including Grimme's D3 van der Waals corrections\cite{grimme2010consistent}. We computed the gyromagnetic ratio of the central spin from the Land\'e-${\bf g}$ tensor, and the ${\bf g}$-tensor and hyperfine coupling ${\bf A}$-tensors (as shown in Eq.~\ref{eq:est_atensor}) using the PBE0 functional\cite{adamo1999toward} with the ZORA-def2-TZVPP basis set. To compute the spin-bath coupling tensor from a density functional theory calculation, we employ the expression
\begin{align}
    \label{eq:est_atensor}
    {\bf A}_{i} ^{ab} =  \frac{\gamma_s \gamma_i}{2S} \hbar^2 \left( \frac{8 \pi}{3}  \rho_s(0) +  \int d {\bf r \rho_s({\bf r}}) \left[ \frac{3 {\bf r}^a {\bf r}^b - | {\bf r}|^2 \delta_{ab}}{|{\bf r}|^5}  \right] \right),
\end{align}
where $\rho_s({\bf r})$ is the electron spin density of the qubit with spin $S$ at a position ${\bf r}$ relative to bath spin $i$. 

For solid-state systems (NV center and $^{31}$P), we employed a point dipole approximation to calculate the coupling tensors (as shown in Eq.~\ref{eq:pd_atensor}) between the central spin and bath spins \cite{Onizhuk2021pycce}. The point-dipole form of the spin-bath coupling tensor is:
\begin{align}
    \label{eq:pd_atensor}
    {\bf A}_i^{ab} = - \gamma_s \gamma_i \frac{\mu_0}{4\pi} \hbar^2 \left[ \frac{3 {\bf r}_i^a  {\bf r}_i^b - | {\bf r}_i|^2 \delta_{ab}}{|{\bf r}_i|^5}  \right]
\end{align}
for a bath spin $i$ located at a distance ${\bf r}_i$ from the qubit. We obtain the gyromagnetic ratios of the central electronic or nuclear spin and the concentrations of bath spins from the EasySpin database via the pyCCE interface. We compute bath-bath interaction tensors ${\bf J}_{i, j}$ from a point dipole approximation for all the systems considered.

\label{SI:parametrization}

\section{SB-tMPS Procedure}

\label{SI:sbtmps-details}

Here, we describe the technical details of the SB-tMPS algorithm from the main text. We start by introducing the construction of the matrix product state (MPS) for the density matrix corresponding to the initial state. Following this, we introduce the construction of the matrix product operator (MPO) for the Liouvillian and describe the advancements we have made here to slash the cost of simulations. We further present a detailed description of the time-dependent variational principle (TDVP) algorithm for updating the MPS and explicitly provide all the steps required to carry out the propagation.

\begin{figure}[!htb]
    \centering
    \includegraphics[scale=0.47, trim={0 3.5cm 0 0}]{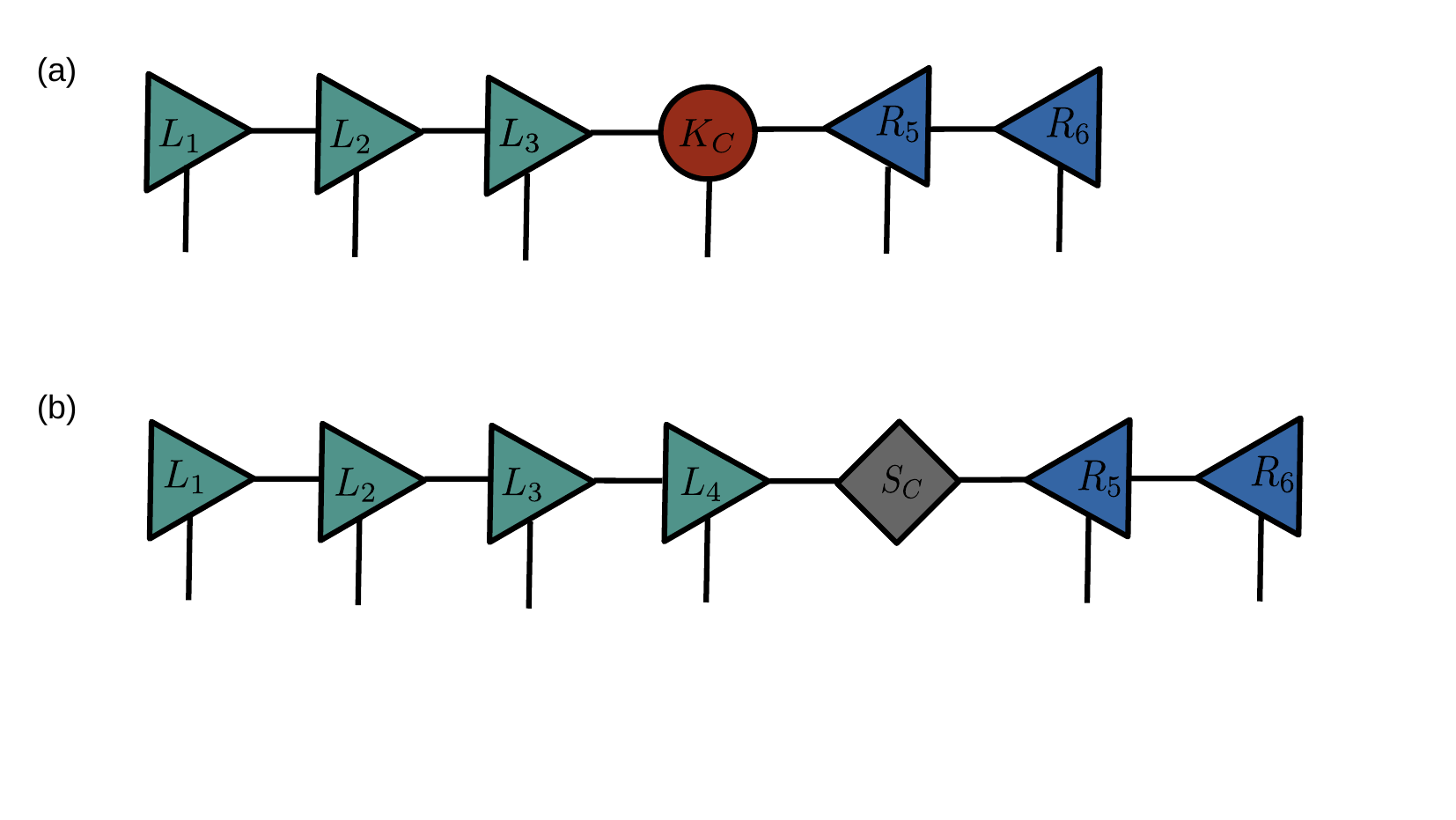}
    \caption{Mixed-Canonical Form of the MPS with a one-site center (a) and a zero-site center (b)}
    \label{fig:mps-orthog}
\end{figure} 

\subsection{Preparation of MPS and MPOs}

We start by adopting the MPS formulation\cite{Schollwock2011AP, Orus2014AP} to express the density matrix as,  
\begin{align}
    &\rho_{i,j,k_1,l_1,\cdots,k_N,l_N} = \sum_{\{r_1,\dots,r_{2N+1}\}} W_1(i,r_1)W_2(r_1,j,r_2) \cdots \nonumber \\
    &\qquad \quad W_{2N+1}(r_{2N},k_N,r_{2N+1})W_{2N+2}(r_{2N+1},l_N)\,
\end{align}
where $i$ and $j$ label the sensor degrees of freedom, while $k_i$ and $l_i$ index the row and column degrees of freedom of the $i$th bath spin. The tensors $W$ are three-dimensional, encoding the correlations between row and column indices and neighboring sites in the MPS representation.  Specifically, we start by preparing a product-state MPS as the initial state : $\rho_0 \otimes \rho_1 ... \otimes \rho_N$ where $\rho_0$ (qubit spin) is prepared in the $|+\rangle$ state and $\rho_1 ...\rho_N$ (bath spins) are infinite-temperature states. 

While the ``split'' MPS representation is popular in quantum  dynamics\cite{guan2024mpsqd,Shi2018JCP,ke2022hierarchical,Jaschke2018QST}, we instead apply the Choi transformation\cite{li2023tangent} (as shown in Fig.~\ref{fig:mps-schem-SI}(b)) to transform to Liouville space, vectorizing the density matrix,
\begin{align}
     |\rho_{i,j,k_1,l_1,\cdots,k_N,l_N}\rangle\rangle = \sum_{\{r_1,...,r_N\}} &\tilde W_1(ij,r_1)\tilde W_2(r_1,k_1l_1,r_2)\cdots \nonumber \\
 &\tilde W_{N+1}(r_N,k_Nl_N).
\end{align}

We prepare a Matrix Product Operator (MPO) by constructing a direct sum of one-site and two-site MPOs to represent the action of $-i\mathcal L \rho$, represented in Liouville space. When constructing the MPO for the spin-bath Hamiltonian, we use a singular value decomposition scheme to truncate the lowest singular values. 

\subsection{Imposing hierarchical SVD thresholds}

The Hamiltonian we use for simulating spin dynamics hosts a wide range of coupling strengths. Since bond dimension grows with increasing connectedness of the Hamiltonian, we impose a hierarchy of SVD truncation thresholds during MPO construction to increase the efficiency of our simulations. While we designed the strategy below with many-spin Hamiltonians in mind, it can be generally applied to any type of Hamiltonian that one is interested in treating with tensor networks. 

We begin by comparing the relative strengths of Hamiltonian coupling terms and use this to inform our choice of SVD threshold for the corresponding term in the MPO. For example, if the maximum intra-bath coupling strength is significantly weaker than the average sensor-bath coupling strength, we apply a stronger SVD truncation to the MPOs corresponding to intra-bath coupling. This can be generalized to any separation of energy scales in the Hamiltonian. If some terms in the Hamiltonian represent strong coupling, with $\lambda^S_{av}$ as their average coupling strength and other terms in the Hamiltonian represent weak coupling, with $\lambda^W_{max}$ being their maximum coupling strength, we use the ratio $\lambda^W_{\max} / \lambda^S_{{\rm av}}$ to set the weak coupling SVD truncation. If the ratio is small, a more aggressive SVD truncation parameter can be adopted for the MPO terms corresponding to weak coupling. The weak coupling SVD threshold is a parameter that can be systematically reduced until convergence is achieved. 

As an illustrative example, in our simulations of the NV-center systems and the BSBS molecular system, we employed a truncation threshold of $r_{\rm{tr}} = 10^{-1}$ for the weak coupling terms (the intra-bath couplings), while for the medium and strong coupling terms (sensor-bath coupling) we retained the standard high-precision truncation of $r_{\mathrm{tr}} = 10^{-14}$, thus balancing computational efficiency and numerical accuracy. 

\subsection{Evolution of the MPS}

The von Neumann equation,
\begin{align}
\label{eq:eom}
     \frac{d}{dt}\rho = -i\mathcal L\rho
\end{align}
dictates the dynamics of the interacting spin bath system, with $\rho$ as the full density matrix and $\mathcal L \cdot = [H_{\rm T},\cdot]$.

\begin{figure}[!b]
    \centering
    \includegraphics[width=\linewidth]{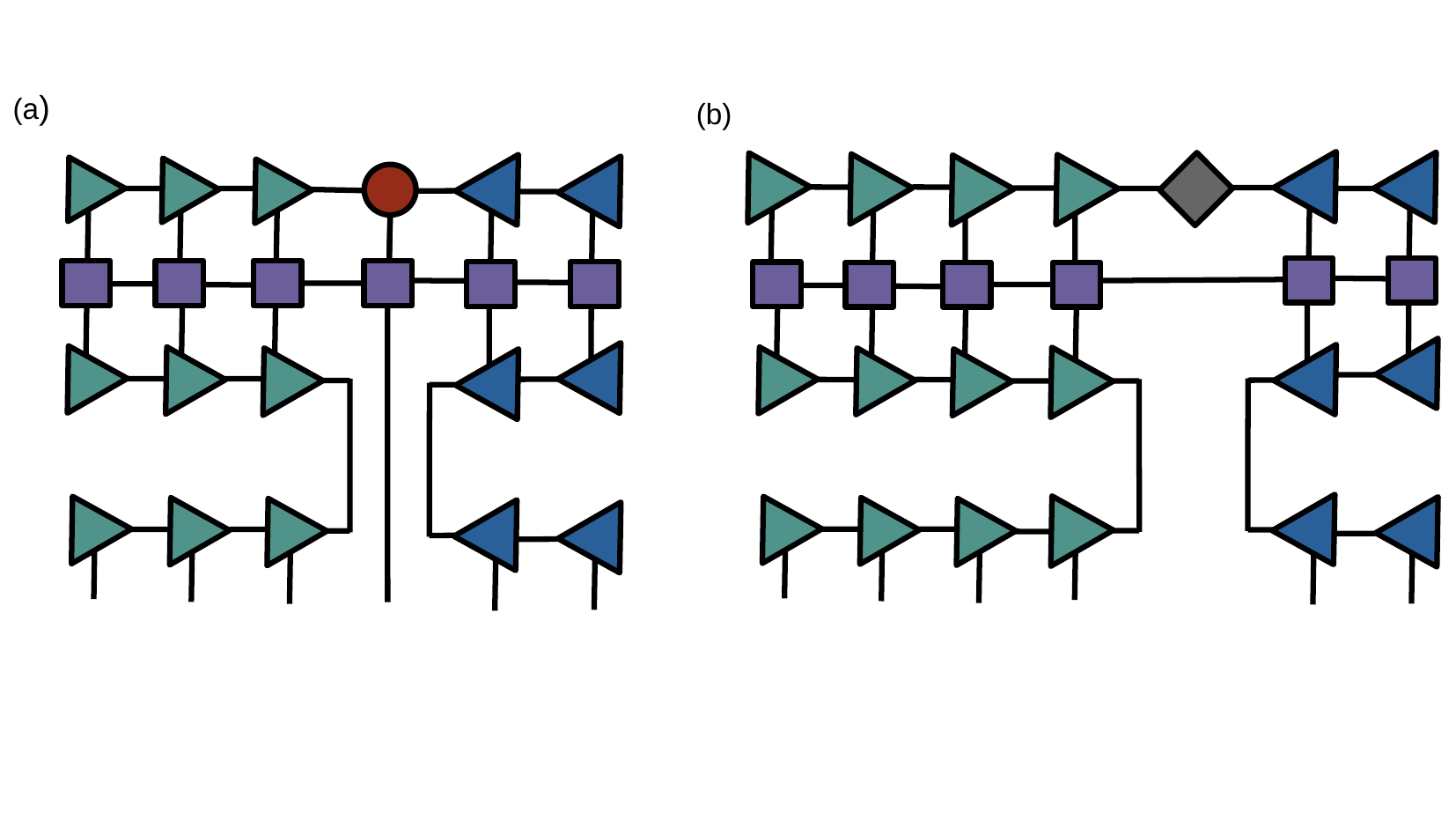}
    \caption{(a) The forward term in RHS of Eq~\ref{eq:projfwd} (b) The backwards term in RHS of Eq~\ref{eq:projback}}
    \label{fig:mps-left}
\end{figure} 

To propagate the density matrix, one can employ the fourth-order Runge–Kutta (RK4) to solve Eq.~\ref{eq:eom} combined with standard tensor truncation procedures to maintain the MPS within a manageable size\cite{oseledets2011tensor}. Although RK4 is stable and accurate when using appropriate truncation parameters\cite{greene2017tensor,guan2024mpsqd}, it can be inefficient. Instead, we adopt the time-dependent variational principle (TDVP)\cite{Haegeman2016PRB} in which the evolution is projected onto the tangent space of the MPS manifold\cite{haegeman2011time,lubich2015time, Haegeman2016PRB,yang2020time,hackl2020geometry}. This effectively fixes the bond dimension of the MPS, eliminating the need to implement a truncation step. This is equivalent to solving the following projected equation:
\begin{equation}
    \frac{d\rho}{dt} = -i \mathcal{P}_{\mathcal{T}, \mathcal{M}_{\rm MPS}} \mathcal{L} \rho.
    \label{eq:TDVP}
\end{equation}

\begin{figure}[!htb]
    \centering
    \includegraphics[width=\linewidth]{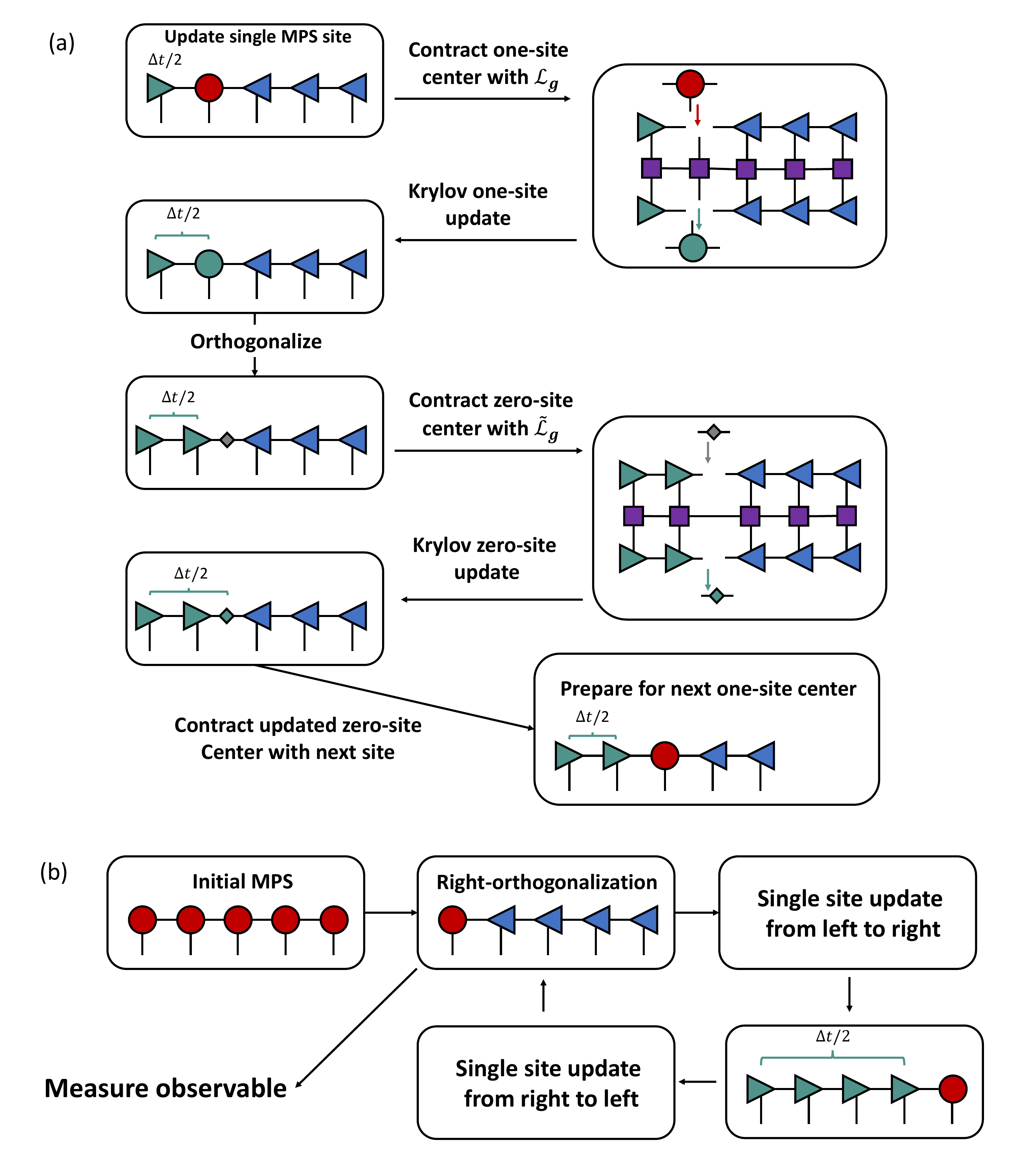}
    \caption{(a) Single-site update procedure. (b) Algorithm for TDVP propagation}
    \label{fig:tdvp-algo}
\end{figure} 

For completeness, we showcase the form of the tangent space projection operator. MPS representations are invariant under the gauge transformation of the site tensors, $\widetilde{W}_i = G_{i-1}^{-1}W_i G_i$. This allows us to bring the MPS into a mixed-canonical form with a one-site center (Eq. ~\ref{eq:onesitecanon}) or a zero-site center (Eq. ~\ref{eq:zerositecanon})---see Fig.~\ref{fig:mps-orthog}. 
\begin{align}
    A(k_1,k_2,\dots,k_N ) &= \sum_{\alpha,\beta} 
    L_{\alpha}^{[1:n-1]}[K_C(n)]_{\alpha,\beta}R_\beta^{[n+1:N]}, \label{eq:onesitecanon}\\
    A(k_1,k_2,\dots,k_N ) &= \sum_{\alpha,\beta} 
    L_{\alpha}^{[1:n]}[S_C(n)]_{\alpha,\beta}R_\beta^{[n+1:N]}\label{eq:zerositecanon},
\end{align}
where $L_\alpha^{[1:n]}$ and $R_\beta^{[n+1:N]}$ denote the left and right orthonormal basis tensors, respectively, with the conditions
\begin{align}
    &\sum_{\sigma} \left(L^{[1:n]}_\alpha\right)^\dagger L^{[1:n]}_\alpha = {\mathbb{I}}, \\
    &\sum_{\sigma} R^{[n+1:N]}_\beta \left(R^{[n+1:N]}_\beta\right)^\dagger = {\mathbb{I}} ,
\end{align}
ensuring that these tensors contract with their complex conjugates to the identity. Accordingly, the tangent-space projection operator can be expressed as\cite{lubich2015time}
\begin{align}
    {\mathcal P}_{\mathcal{T}, \mathcal{M}_{\rm MPS}} 
    = \sum_{n=1}^N  P_L^{[1:n-1]} \otimes {\mathbb{I}}_n \otimes P_R^{[n+1:N]}
    - \sum_{n=1}^{N-1} P_L^{[1:n]} \otimes P_R^{[n+1:N]},
    \label{eq:projection-operator}
\end{align}
where,
\begin{align}
    P_L^{[1:n]} = \sum_{\alpha} L_\alpha^{[1:n]} {L_\alpha^{[1:n]}}^T, 
    \qquad 
    P_R^{[n:N]} = \sum_{\beta} R_\beta^{[n:N]} {R_\beta^{[n:N]}}^T,
\end{align}
and Eq.~\ref{eq:TDVP} can be rewritten into two terms, 
shown in Fig.~\ref{fig:mps-left}.

To efficiently solve Eq.~\ref{eq:TDVP}, we employ a Lie–Trotter decomposition of the projection operator, which allows us to propagate the MPS forward with $N$ equations of the form:
\begin{align}
    \frac{d \rho}{dt} = -i P_L^{[1:n-1]} \otimes {\mathbb{I}}_n \otimes P_R^{[n+1:N]} \mathcal{L} \rho,
    \label{eq:projfwd}
\end{align}
and backwards with $N-1$ equations of the form: 
\begin{align}
   \frac{d \rho}{dt} = i P_L^{[1:n]} \otimes P_R^{[n+1:N]} \mathcal{L} \rho.
   \label{eq:projback}
\end{align}
Although solving Eq.~\ref{eq:TDVP} with the projection operator in Eq.~\ref{eq:projection-operator} yields an approximate result, the associated error remains controllable and can be systematically reduced by increasing the MPS bond dimension and decreasing the timestep.

We implement the update for a single site over a time step $\Delta t$ via left-to-right and right-to-left sweeps~\cite{lubich2015time}. During each sweep, we perform a sequence of orthogonalization operations, evaluate the generalized Liouvillian at site $n$, and update the one-site centered site $K_C$ together with the zero-site centered site $S_C$ according to
\begin{align}
    K_C(n,t+\Delta t) &= \exp[-i\mathcal{L}_{\rm g}(n)\Delta t]\,K_C(n,t), \\
    S_C(n,t+\Delta t) &= \exp[i\tilde{\mathcal{L}}_{\rm g}(n)\Delta t]\,S_C(n,t),
\end{align}
where the definitions of the generalized Liouvillians are given in Ref.~\onlinecite{haegeman2011time}. The time evolution governed by these equations is solved within a Krylov subspace using the iterative Arnoldi procedure\cite{golub2013matrix, saad2003iterative}.

The entire detailed evolution procedure for TDVP is illustrated in Fig.~\ref{fig:tdvp-algo}.

\subsection{Measuring the reduced density matrix}

Our observables of interest for the many-spin system are the coherence and population dynamics (as shown in Fig.~\ref{fig:mps-schem-SI}(c)). One can obtain these from the diagonal and off-diagonal terms of the reduced density matrix. We extract the reduced density matrix for the sensor spin by sequentially contracting successive sites and then tracing out the extra bonds that result from the contraction. 

\section{Review of CCE}

In CCE\cite{Yang2008CCE1, Yang2009CCE2}, the coherence function is expressed as a product of irreducible terms from bath spin subclusters:
\begin{align}
    \label{eq:cce-prod-exp}
    C(t) = \prod_A \tilde{C}_A(t) = \prod_i \tilde{C}_{\{i\}}(t) \prod_{i,j} \tilde{C}_{\{ij\}}(t)...,
\end{align}
where $\{i,j,\cdots\}$ denote all possible subclusters of the total spin bath that includes spins indexed by $i, j, \cdots$. That is, $\{i\}$ denotes spin subclusters consisting of one spin, $\{ij\}$ denotes spin subclusters consisting of two spins, and so on. Here, each cluster contribution is given by
\begin{align}
    \tilde{C}_A(t) = \frac{C_A(t)}{\prod\limits_{B\subset A} \tilde{C}_B(t)}.
\end{align}
The coherence function for clusters is computed by evaluating the following expression: 
\begin{align}
    C_A(t) = \langle 0 | \hat{U}_A(t) \hat{\rho}(0) \hat{U}_A^{\dagger}(t) | 1 \rangle,
\end{align}
where the time-propagator for the $A$ cluster (with sub-Hamiltonian $H_A$) subject to a set of $N$ CPMG pulses, separated by time delay $\tau$ and characterized by rotational angle $\phi$, is given by:
\begin{equation}
    \label{eq:timeprop}
    \hat{U}_A = \mathcal{T} \left[ {\rm e}^{-\frac{i}{\hbar} \hat{H}_A \tau} {\rm e}^{-\frac{i}{\hbar} \hat{\sigma}_y \frac{\phi}{2}} {\rm e}^{-\frac{i}{\hbar} \hat{H}_A \tau}\right]^N.
\end{equation}

In the original CCE formulation, the cluster Hamiltonian $\hat{H}_A$ is obtained by projecting the Hamiltonian on each central spin state $\ket{0}$ and $\ket{1}$ and tracing out bath spins not included in the cluster. In the generalized CCE (gCCE) formulation, the cluster Hamiltonian includes the full central spin Hamiltonian for computations performed on individual clusters\cite{yang2020longitudinal, Onizhuk2021pycce}. In the gCCE formulation, instead of expanding the coherence function as a product of irreducible cluster terms as in Eq.~\ref{eq:cce-prod-exp}, one expands the reduced density matrix for the central spin\cite{Onizhuk2021PRXQ}. This enables gCCE to access population dynamics, while the conventional CCE formulation can only probe coherence dynamics.

\begin{figure}[t!]
    \centering
    \includegraphics[width=0.4\linewidth]{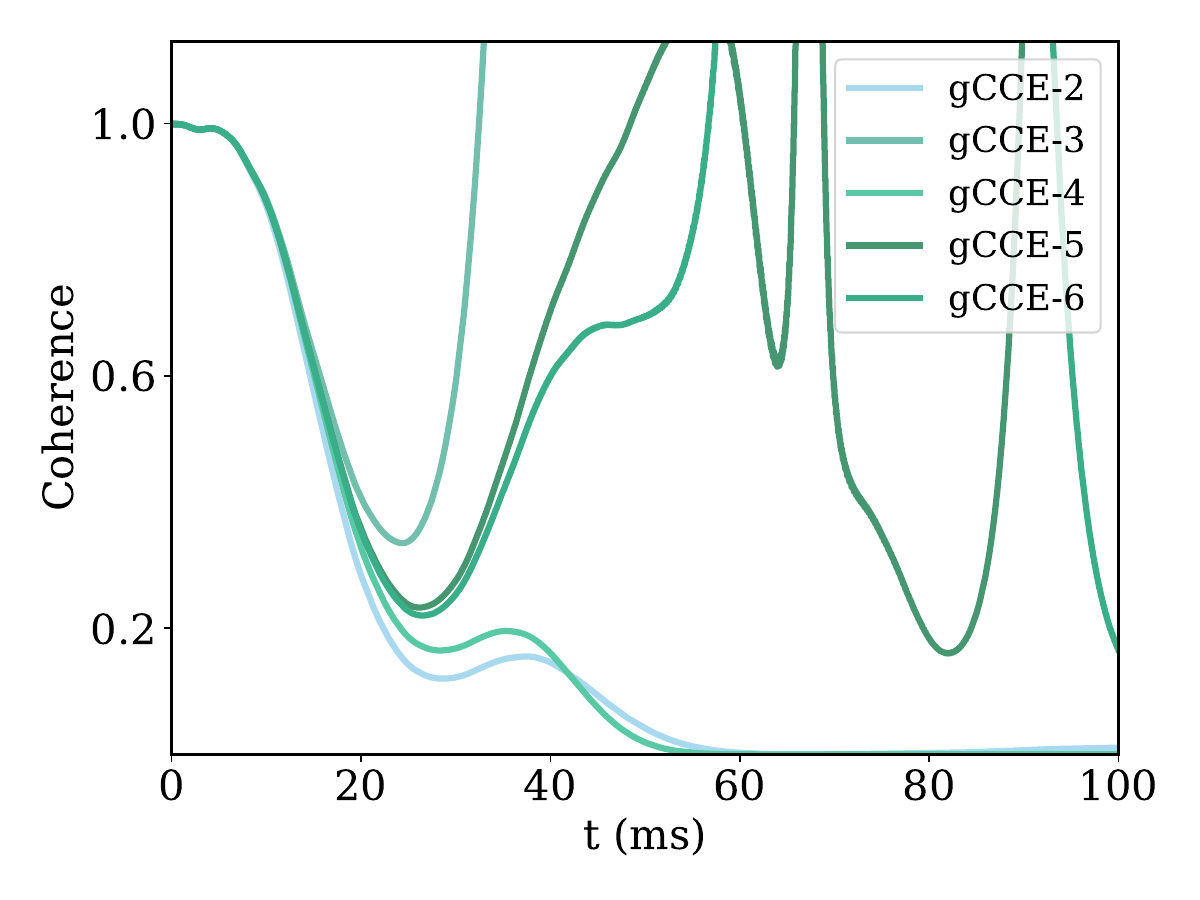}
    \caption{Non-monotonic convergence of the CCE. Spin-Echo dynamics in the $^{31}$P nuclear spin qubit at various CCE orders.}
    \label{fig:SI-non-monotonic-convergence}
\end{figure} 

\label{SI:CCE}

\subsection{Efficiency and convergence}

Like perturbation theory, the CCE method is most efficient when the dynamics converges at a low cluster order. In such cases, the total computational effort required to evaluate the dynamics via CCE is lower than that associated with exactly solving the full many-body problem. However, when a low-order expansion is insufficient, CCE can become computationally prohibitive. By construction, when taking the CCE calculation to order its maximum $K$, i.e., the total number of spins in the Hamiltonian, CCE yields the exact result. Because all lower orders have been computed, the CCE computation becomes more expensive than the one-shot, exact calculation. In all CCE simulations, results must be converged with respect to the order of its expansion, although---as we show in Fig.~\ref{fig:SI-non-monotonic-convergence}---its convergence is non-monotonic, making it difficult to assess when it is reached.

Because the cost of CCE rises with system size and the order to which it is taken, approaches have been developed to enhance its efficiency. For example, it is common to implement a distance cutoff (${\rm r_{dip}}$), which restricts the maximum distance up to which two bath spins can be considered part of the same cluster. If this option is used, the dynamics must also be converged with respect to this parameter, although systematic convergence has not been proven. 

\begin{figure}[b]
    \centering
    \includegraphics[width=\linewidth]{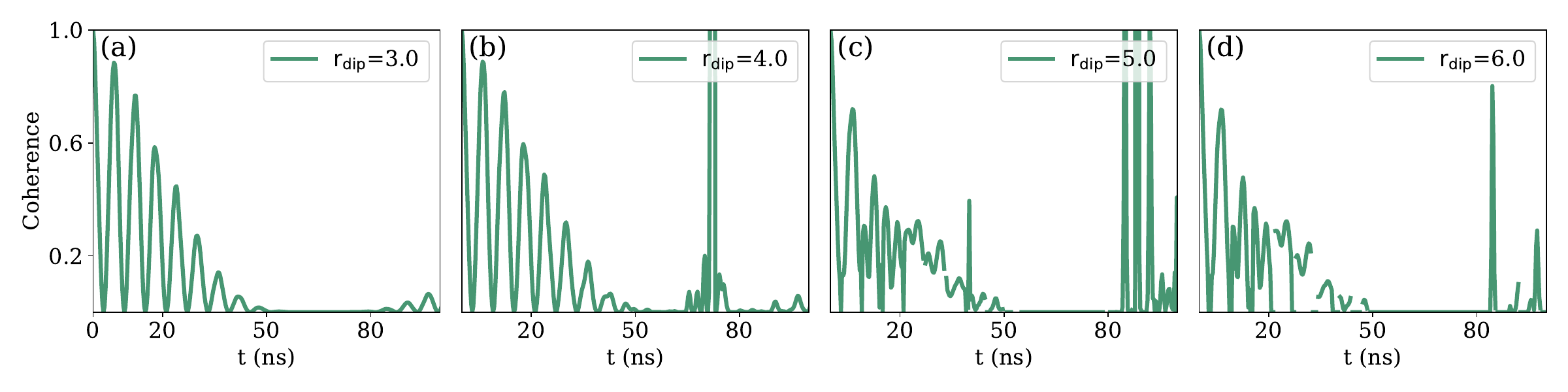}
    \caption{Increasing ${\rm r_{dip}}$ in CCE for the molecular BSBS-2Et system does not allow for convergence. Qualitatively the dynamics changes as ${\rm r_{dip}}$ is increased, however the number of numerical instabilities also increases. Further, a large amount of data is missing, owing to divisions by zero, which does not allow for comparison between successive increases in ${\rm r_{dip}}$.}
    \label{fig:NaN}
\end{figure} 

\begin{figure}[t]
    \centering
    \includegraphics[width=0.4\linewidth]{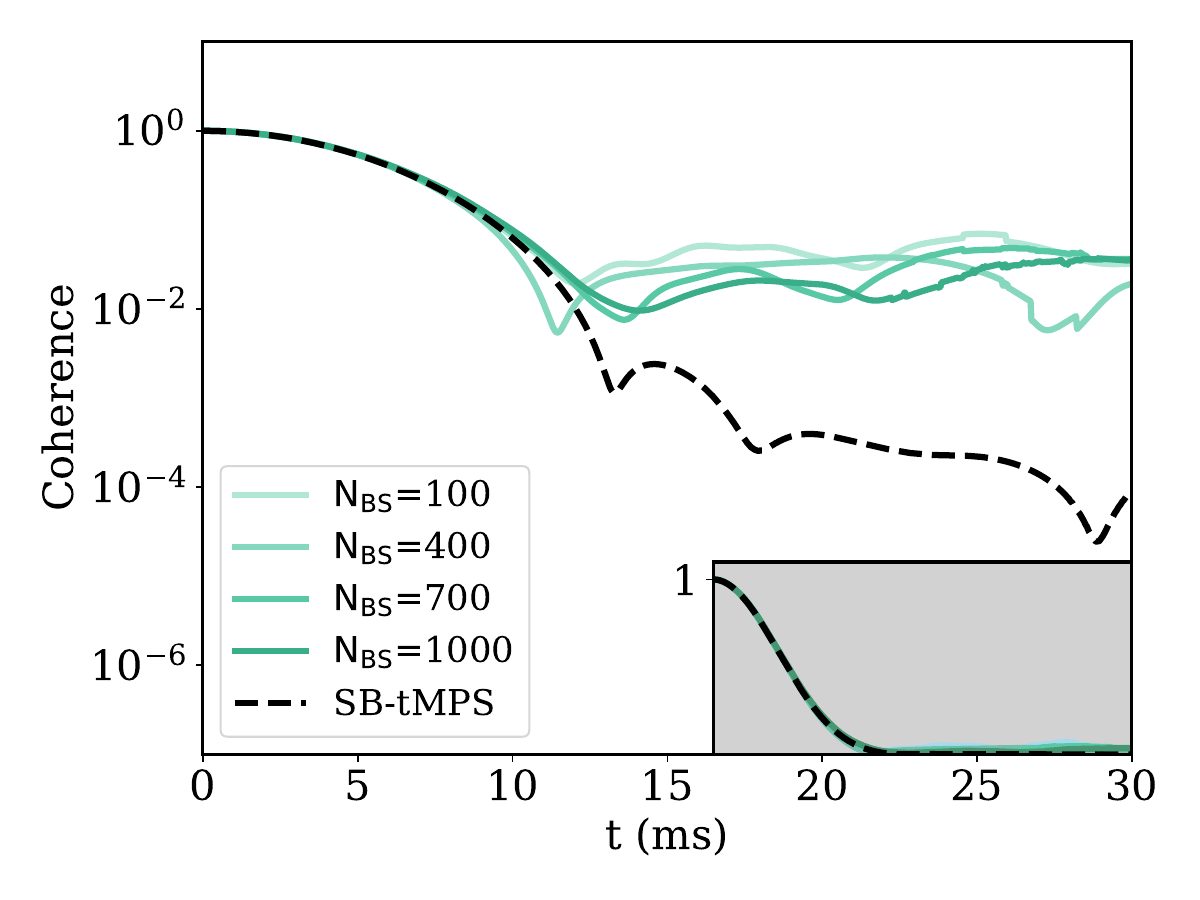}
    \caption{Bath sampling and its inability to recover detailed long-time structure. Disagreement between our numerically exact MPS dynamics for the $^{31}$P nuclear spin qubit under free induction decay and gCCE-4 dynamics obtained with increasing degrees of bath sampling.}
    \label{fig:SI-bath-sampling-P}
\end{figure} 

\subsection{Instabilities}

Although convenient for pure dephasing-type problems, especially in impurity-type Hamiltonians, the product formulation of CCE lends itself to numerical instabilities that can arise from dividing by small cluster contributions. In Fig.~\ref{fig:NaN}, we demonstrate this by increasing ${\rm r_{dip}}$. These numerical instabilities can make it difficult to systematically improve the accuracy of some CCE calculations with respect to their convergence parameters. 

As mentioned in the main manuscript, techniques to ameliorate these instabilities have been developed and implemented, such as Monte Carlo sampling of bath states\cite{Witzel2012, Onizhuk2021pycce, Onizhuk2021PRXQ}. In this technique, one constructs several ``pure bath'' states, each having a bath spin initialized in the spin-up or spin-down state. One then evaluates the coherence dynamics for each of these configurations using CCE and the results are combined as such: $C_{\rm BS}(t) = \sum_i p_i C_i(t)$, where $C_i(t)$ is the coherence evaluated from the sampled pure state $i$ and $p_i$ is the probability to sample the state $i$. For a bath at infinite temperatures, all probabilities $p_i$ are equal. We apply this technique on the system considered in Fig.~\ref{fig:molecule}~(b) using 100-1000 sampled bath states as well as in the nuclear spin qubit as shown in Fig.~\ref{fig:SI-bath-sampling-P}. In both examples, it is clear that pure state bath sampling only superficially removes spikes in the results that go beyond their physical limits (i.e., 0 and 1), but the results do not exhibit greater accuracy. 

\begin{figure}[b]
    \centering
    \includegraphics[width=0.4\linewidth]{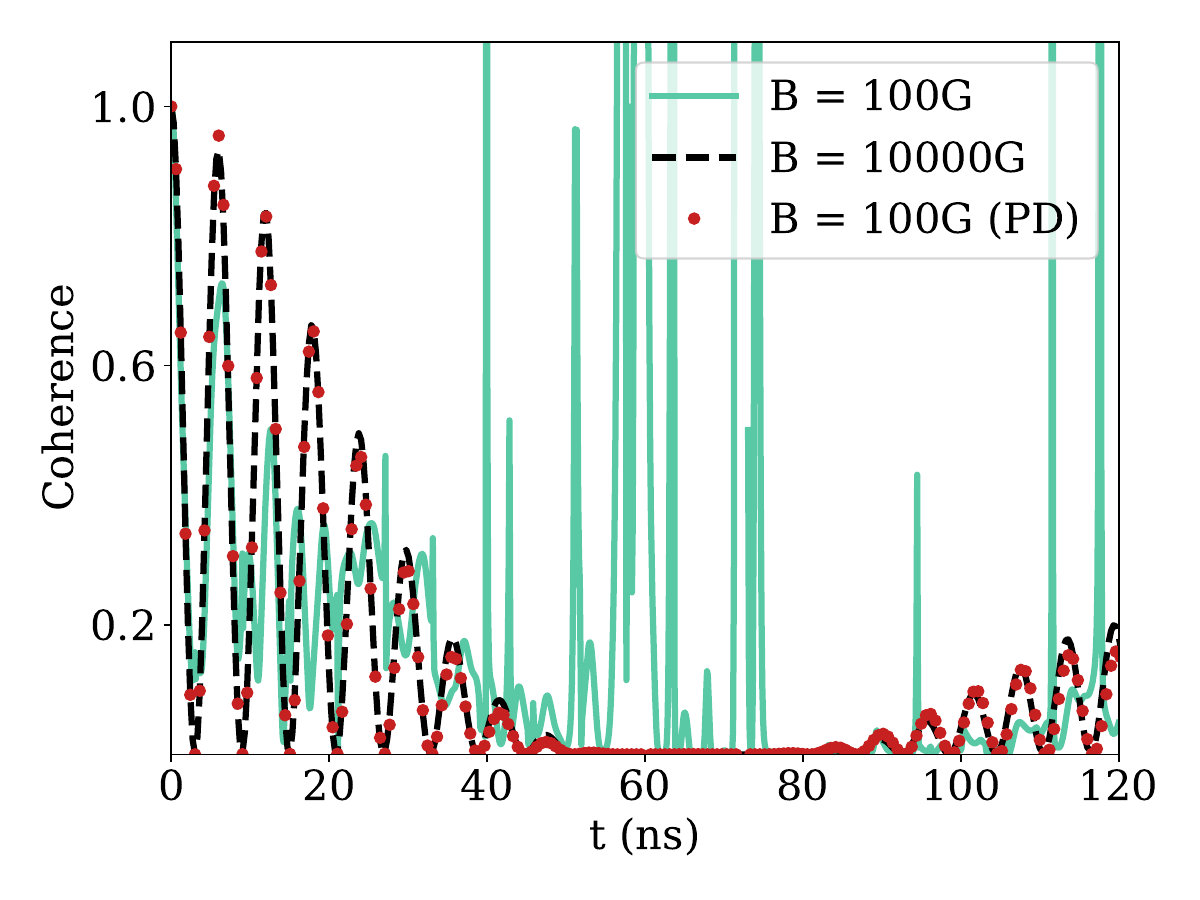}
    \caption{The gCCE dynamics of the BSBS-2Et molecule have no numerical instabilities in the pure dephasing limit. We get to this limit by either forcing the system into the pure dephasing (PD) limit by explicitly turning off $k \neq z$ couplings in Eq.~\ref{eq:sbham} (red) or by increasing the magnetic field sufficiently (black).}
    \label{fig:cce-pd-mol}
\end{figure}

While the gCCE formulation enables us to simulate population dynamics, we find that the method struggles to produce accurate dynamics outside of the pure dephasing regime, as we show in Fig.~\ref{fig:molecule}. We demonstrate that the cause of these instabilities is indeed the system experiencing population relaxation effects by pushing the system into the pure dephasing (PD) limit in two ways. The first is by increasing the magnetic field, which minimizes the impact of relaxation effects to the point where they become negligible. The second approach is to explicitly set to zero the $k \neq z$ interactions in Eq.~\ref{eq:sbham}, thereby forcing the system into the pure dephasing limit. In both these cases, we observe that gCCE produces numerically stable dynamics, as shown in Fig.~\ref{fig:cce-pd-mol}.

\begin{figure}[t!]
    \centering
    \includegraphics[width=\linewidth]{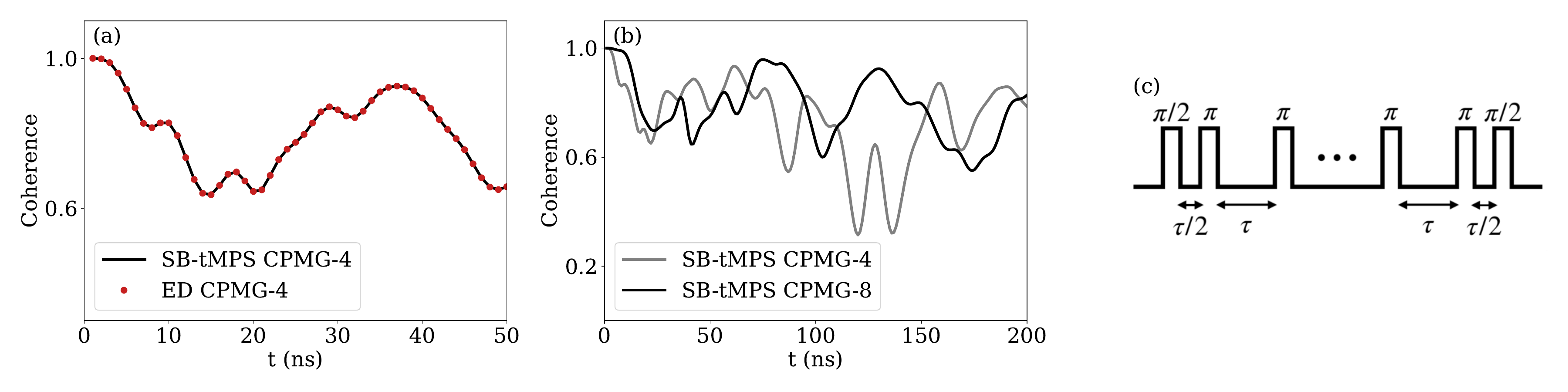}
    \caption{(a) CPMG-4 dynamics for the unsubstituted BSBS molecule with ED and SB-tMPS (b) Many-pulse coherence dynamics of the BSBS-2Et system using SB-tMPS (c) Schematic of the CPMG pulse sequence.}
    \label{fig:mps-cpmg}
\end{figure} 

\section{SB-tMPS Coherence dynamics under CPMG pulse sequences}
\label{SI:pulsed}

Our SB-tMPS method enables us to simulate the dynamics of spin qubits under arbitrary pulse sequences. Here, we outline the strategy that we employ to parallelize dynamics under Carr-Purcell-Meiboom-Gill\cite{carr1954effects, meiboom1958modified} (CPMG) pulse sequences (see Fig.~\ref{fig:mps-cpmg}(c)). In the simulation of CPMG pulse sequences, we define the time interval between the initial $\pi/2$ pulse and the first $\pi$ pulse as $\tau$. The total simulation time is then given by $2\tau \times N_p$, where $N_p$ denotes the number of $\pi$ pulses. To obtain the coherence dynamics, we perform $N_s$ simulations in parallel by scanning over different values of $\tau$, yielding $N_s$ data points in total. In practice, the $\pi/2$ and $\pi$ pulses are implemented by applying the corresponding spin-rotation operators directly to the system site of the SB-tMPS, ensuring an accurate representation of pulse actions within the MPS framework. We benchmark our method for pulsed dynamics against ED for the unsubstituted BSBS molecule consisting of 8 $^{1}$H bath spins and 2 $^{77}$Se (see Fig.~\ref{fig:mps-cpmg}(a)). We then simulate the CPMG-4 and CPMG-8 dynamics for the BSBS-2Et molecule in Fig.~\ref{fig:mps-cpmg}(b).

\section{CCE and SB-tMPS runtime scaling}
\label{SI:CCEruntime}

\begin{figure}[b]
    \centering
    \includegraphics[width=\linewidth]{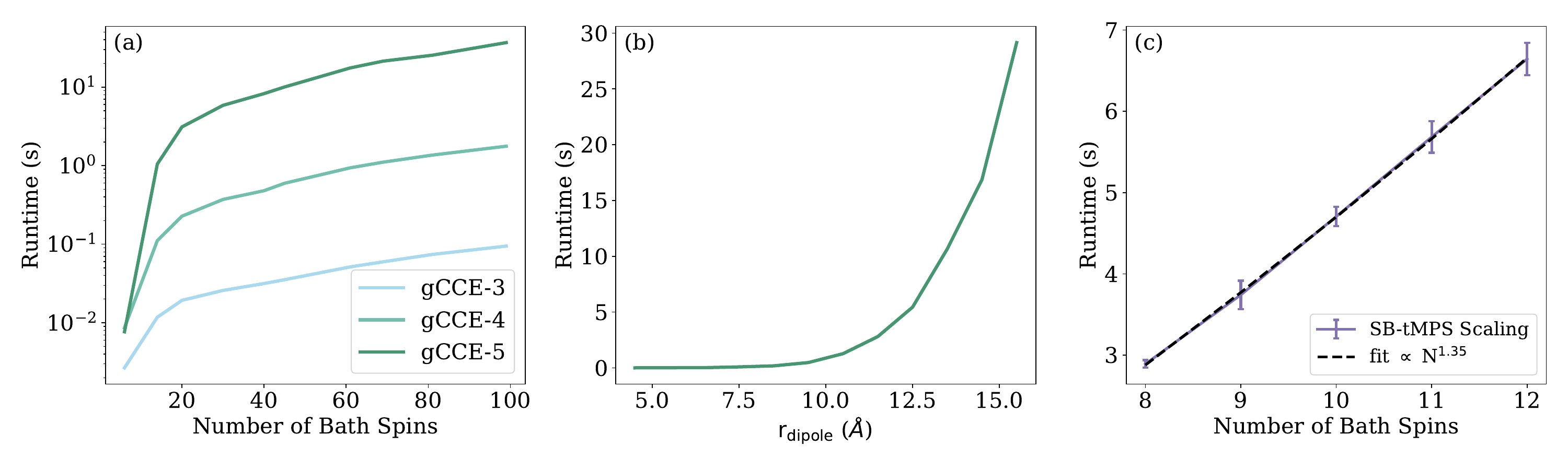}
    \caption{Scaling of gCCE runtime for 10 steps at various orders as a function of the number of bath spins (a), and as a function of ${\rm r_{dipole}}$ (b). (c) Scaling of SB-tMPS runtime for 10 steps with number of bath spins for the for the $^{31}$P nuclear spin qubit system.}
    \label{fig:cce-scal}
\end{figure} 

Now we turn to a rigorous testing of the scaling of both the CCE and SB-tMPS methods. In Fig.~\ref{fig:cce-scal}(a) and (b), we show that the CCE method scales exponentially with both cluster order and the ${\rm r_{dip}}$ parameters, respectively. This makes it computationally expensive to converge CCE parameters systematically for large, strongly-interacting spin systems.

In Fig.~\ref{fig:NV}(c), we showcase that our SB-tMPS approach scales linearly with the number of bath spins for the NV center, where qubit-bath interactions are significantly stronger than intra-bath interactions, allowing us to use an aggressive SVD truncation ($r_{tr} = 10^{-1})$. For the $^{31}$P nuclear spin qubit, where the qubit-bath interactions are of similar magnitude as intra-bath interactions, we observe modest supra-linear scaling ($\sim N^{1.35}$) with the bath size---see Fig.~\ref{fig:cce-scal}(c). Thus, even in cases where our hierarchical SVD truncation method does not give significant savings, our SB-tMPS shows more favorable scaling than CCE. 

\clearpage
\bibliography{main}